\documentclass{article}[14pt]
\usepackage{graphicx}
\usepackage{amsmath}
\usepackage{amssymb}
\usepackage{braket}
\usepackage{longtable}
\usepackage{mathtools}
\usepackage{hyperref}
\usepackage{breqn}
\usepackage{cite}

\usepackage{ifthen}
\usepackage{calc}
\usepackage{bm}
\usepackage{color}
\usepackage{psfrag}
\usepackage{wrapfig}
\usepackage{subfigure}
\usepackage{rotating}
\usepackage{tikz}
\usetikzlibrary{shapes.arrows}
\usetikzlibrary{decorations.pathmorphing}\usepackage{mdframed}
\usepackage{braket}
\usepackage{breqn}
\usepackage{upgreek}
\usepackage{cleveref}
\numberwithin{equation}{section}
\newcommand\email[4]{#1@#2.#3.#4}

\newcommand{\be}{\begin{equation}}
	\newcommand{\bs}{\begin{split}}
		\newcommand{\ben}{\begin{equation}\nonumber}
			\newcommand{\ee}{\end{equation}}
		\newcommand{\es}{\end{split}}
	\newcommand{\bi}{ \begin{itemize} }
		\newcommand{\ei}{ \end{itemize} }

\begin{document}
		
		\title{Application of analytic functionals to mean field theory and Wilson- Fisher fixed point }
		\author{
			Bhaskar Jyoti Khanikar\thanks{\email{bhaskar70707}{gmail}{com}{}}
			~and~
			Subir Mukhopadhyay\thanks{\email{subirkm}{gmail}{com}{}}
		}
		\date{
			{\small Department of Physics, Sikkim University, 6th Mile, Gangtok 737102}
		}
		
		\maketitle
		
		\begin{abstract}
			\small
			\noindent We consider application of the analytic functional approach to the conformal field theories associated with mean field theory and Wilson-Fisher fixed point. 
			We study the constraints imposed by the crossing symmetry on the coefficients of the operator product expansion. Making use of these constraints along with 
			a few other additional inputs, we obtain expressions of the coefficients of the operator product expansion up to first order of $\epsilon$.
			
		\end{abstract}
		
	\clearpage

		\section{Introduction}
		
		In recent years the conformal bootstrap program is undergoing a rapid development. The numerical bootstrap, making use of unitarity and crossing symmetry has grown into a very powerful machinery. It has successfully put general constraints on the space of conformal field theories (CFT) and determine the data of specific models with unprecedented accuracy \cite{Poland:2018epd,Rychkov:2023wsd}.  Our understanding of various aspects of the bootstrap constraints from an analytical perspective is also developing at a fast pace \cite{Bissi:2022mrs}. 
		
		At present there are several avenues through which analytic understanding of the bootstrap is evolving. For $d>2$, in the limit of large spin, it has been shown \cite{Fitzpatrick_2013} that operator product expansion (OPE) of a scalar primary with itself will involve an infinite tower of operators with twists $\tau\rightarrow 2\Delta_\phi + 2 n$, where $n$ are non-negative integers and the anomalous dimension $\gamma(n,\ell)$ falls off as $\frac{\gamma_n}{\ell^{\tau_m}}$, where $\tau_m$ is the minimal twist. Related predictions can be made about the OPE coefficients as well. In \cite{Carmi:2019cub}, using an analogy with Kramers-Kronig relation in optics, four-points of scalar operators has been shown to be expressed in terms of an integral transform of  its absorptive part, identified as a double discontinuity of the correlator, where the kernel of the integral transform is independent of the theory and is determined by resumming the data obtained by the Lorentzian inversion formula. Lorentzian inversion formula \cite{Caron-Huot:2017vep,Simmons-Duffin:2017nub,Kravchuk:2018htv} is the related result where in a given OPE channel the coefficient function of partial wave expansion can be expressed in terms of the double discontinuities of other two channels.
		
		In another approach, the exact bootstrap bounds for CFTs in one dimensions was studied \cite{ mazac2016analytic, mazavc2019analytic, mazac2018analytic}. These analytic results can be reinterpreted in the context of modular bootstrap and sphere packing problem in Euclidean geometry \cite{Hartman:2019pcd}. This has further been extended to construction of analytic functionals in the context of four point functions in CFT in one dimensions \cite{Mazac:2018qmi} and two point functions in the context of boundary conformal field theory \cite{Kaviraj:2018tfd, Mazac:2018biw} in any dimension. The analytic functional for four-point functions in conformal field theories of dimensions greater than 1 are developed in \cite{mazavc2021basis}. Action of these functionals are presented as double contour integral with suitable kernels. The relation between this approach and the aforementioned dispersive method has also been pointed out \cite{Carmi:2019cub}. 
		
		Yet another procedure uses a close counterpart of  momentum space S-matrix in conformal field theory, which is essentially a Mellin transform of the four point correlator \cite{Gopakumar:2016wkt,Gopakumar:2016cpb}. Combining this representation with observation of Polyakov, one obtains constraint equation on the residue of the poles of the Mellin amplitude. The validity of the Mellin representation has been studied in \cite{Penedones:2019tng} and they obtained non-perturbative dispersion relations in the Mellin space. They have also obtained interesting sum rules for CFT data from the absence of exact double-twist quantum numbers in the spectrum. 
		
		In this paper, we will be concerned with the second method involving analytic functionals. A complete basis of such analytic functionals are constructed in general dimension \cite{mazavc2021basis}. Applying these functionals on correlators of CFTs leads to strong constraints on CFT data, which may be interesting to study in concrete models. A study has already been done for conformal field theory on real projective space in \cite{Giombi:2020xah}. Here, we have  considered applications to mean field theories \cite{Fitzpatrick_2013,Fitzpatrick:2011dm,Bissi:2022mrs,paulos2020analytic,Bissi_2020,mazavc2021basis} and  perturbative expansions around the mean field theory, known as the Wilson-Fisher fixed point \cite{Giombi:2020xah,Rychkov:2015naa, Basu:2015gpa, Raju:2015fza, Gliozzi:2016ysv, Dey:2016mcs, Roumpedakis:2016qcg, Alday:2016jfr, Gopakumar:2016wkt, Gopakumar:2016cpb, Gliozzi:2017hni, Liendo:2017wsn, Dey:2017fab, Gliozzi:2017gzh,Alday:2017zzv, Gopakumar:2018xqi} in $4-\epsilon$ dimensions. Comparing with the exchange Witten diagrams, we obtain relations, which with a few additional assumptions, can be used to obtain  the OPE coefficients and the corrections in them  upto first order. Unlike the case of projective space \cite{Giombi:2020xah} we need to deal with two different indices, which makes the treatment more difficult.
		
		The plan of the paper is as follows. In the next section we will discuss the four point correlators, their expansion in terms of basis, the dual analytic functionals and the relations between the analytic functionals and exchange Witten diagrams. In third section we discuss applications of the analytic functionals on the crossing symmetry equations  in the case of mean field theory and the Wilson-Fisher fixed point. Finally we conclude in section 4. Some of the calculations involving the recursion relations are moved to an appendix.
		
		\section{Analytic Functionals }
		
	\subsection{Preliminaries}
		
		We begin our discussion by reviewing the analytic structure of the generic conformal blocks. In conformal field theory, one of the important concept is the operator product expansion, which enables one to 
		express the product of two operators at close points  in terms of a set of operators. In particular, the operator product  for two scalar primary operators $\phi(x_1)$ and  $\phi(x_2)$ it is given by
		\be \phi(x_1) \phi(x_2) = \frac{       \delta_{12}         } { (x_1-x_2)^{2\triangle_{\phi_1} }}  + \sum\limits_{\mathcal O} C_{12\mathcal{O}} D[x_1-x_2, \partial_{x_2}]\mathcal{O}(x_2)\ee
		where on the right hand side one sums over the primary operators, $C_{12\mathcal{O}}$ are called the OPE coefficients and  $D[x_1,x_2]$ represents a differential operator, whose expression is fixed by the conformal invariance. 
		The scaling dimension of the operators satisfy unitarity bounds for a unitary theory,
		\ben \triangle \geq \Big\{ \begin{array}{cc}\frac{d-2}{2},& \ell = 0 \\ d - 2 + \ell & \ell > 0 \end{array}\ee
		where, $\ell$ denotes the spin of the operator.
		
		In a conformal field theory, the two and three point functions are determined by the conformal symmetry, while the four point functions is determined upto a function of cross ratios,  given by
		\be u = \frac{x_{12}^2 x_{34}^2}{x_{13}^2 x_{24}^2} \quad, \quad v = \frac{x_{14}^2 x_{23}^2}{x_{13}^2 x_{24}^2} .\ee 
		Sometimes it is convenient to switch over another set of variables\cite{dolan2004conformal,dolan2001conformal}, $z$ and $\bar{z}$, which are related to $u$ and $v$ as
		\be u = z\bar{z} , \quad v = (1-z)(1-\bar{z}). \ee 
		In the Euclidean signature, $(z,\bar{z})$ are complex conjugate of each other while in Lorenzian signature, they are independent.  We will be using $(u,v)$ and $(z,\bar{z})$ interchangeably.
		
		The correlation function of four scalar primary operators $\phi$ with scaling dimensions $\triangle_i$, $i=1,..,4$  is given by
		\be \langle \phi(x_1)\phi(x_2)\phi(x_3)\phi(x_4) \rangle = \frac{1}{    (x_{12}^2)^{\frac{\Delta_1+\Delta_2}{2} }      (x_{34}^2)^{\frac{\Delta_3 + \Delta_4}{2} }      }   \left(\frac{  x_{24}^2 }{  x_{14}^2  }  \right)^{\frac{\Delta_{12}}{2}}  \left(\frac{  x_{14}^2 }{  x_{13}^2  }  \right)^{\frac{\Delta_{34}}{2}}  \tilde{\mathcal{G}}(u,v)\label{4point},\ee where we use $\Delta_{ij}=(\Delta_i - \Delta_j)$ and $\tilde{\mathcal{G}}(u,v)$ is termed as conformal partial wave (CPW) in \cite{dolan2001conformal}.
		
		Using the operator product expansion, one can reduce the four-point correlators as a product of two three point correlators involving the operator 
		${\mathcal O}$ and summed over it as,
		\be \label{expand} \langle \phi(x_1)\phi(x_2)\phi(x_3)\phi(x_4) \rangle =  \frac{1}{    (x_{12}^2)^{\frac{\Delta_1+\Delta_2}{2} }      (x_{34}^2)^{\frac{\Delta_3 + \Delta_4}{2} }      }   \left(\frac{  x_{24}^2 }{  x_{14}^2  }  \right)^{\frac{\Delta_{12}}{2}}  \left(\frac{  x_{14}^2 }{  x_{13}^2  }  \right)^{\frac{\Delta_{34}}{2}}  \sum\limits_{\mathcal{O}}    \lambda_{12\mathcal{O}}   \lambda_{34\mathcal{O}}   g^{\Delta_{12},\Delta_{34}}_{   \Delta_{\mathcal O}, \ell_{\mathcal O}   },\ee
		where $\lambda_{12\mathcal{O}}$ and $\lambda_{34\mathcal{O}}$ are the operator product expansion coefficients and thus, the four point correlator reduces into a sum over contributions from exchange operators, denoted by
		$g^{\Delta_{12},\Delta_{34}}_{   \Delta_{\mathcal O}, \ell_{\mathcal O}   }$, called conformal blocks, representing the contribution for a primary operator ${\mathcal O}$ of scaling dimension $\Delta_{\mathcal O}$ and spin $\ell_{\mathcal O}$, and its descendants. 
		
		The conformal blocks are eigenfunctions of the quadratic Casimir operator of the conformal group. Their expressions for even dimensions has been obtained in  \cite{dolan2001conformal}. In particular, for $d=4$, it can be written as
		\be \label{conformalblock} g^{\Delta_{12},\Delta_{34}}_{   \Delta, \ell   } = \frac{1}{(-2)^\ell}\frac{z \bar{z}}{z-\bar{z}} [ k_{\Delta+\ell}(z) k_{\Delta-\ell-2}({\bar z}) - ( z \leftrightarrow \bar z) ],\ee
		where $k_\beta(x) = x^{\beta/2}~  _2F _1 (\frac{\beta - \Delta_{12}}{2}, \frac{\beta + \Delta_{34}}{2} , \beta ; x )$. 
		
		For the sake of simplicity, we restrict  to the scalar operators having equal dimension, {\it i.e} $\Delta_i =\Delta_\phi$ and since $\Delta_{12}=\Delta_{34}=0$ we will write $g_{   \Delta, \ell   }$ for the conformal block. Further, we will change the notation slightly, and redefine the conformal partial wave and the conformal blocks  as \cite{mazavc2019analytic,mazac2018analytic}
		\be \label{redefine} {\mathcal G} = (z\bar{z})^{-\Delta_\phi} \tilde{{\mathcal G}},\quad\quad G_{\Delta, \ell} = (z\bar{z})^{-\Delta_\phi} g_{\Delta, \ell} .\ee  
		
		The crossing symmetry involves the relation between OPE in terms of $\phi(x_1), \phi(x_2)$ and $\phi(x_3), \phi(x_4)$, and OPE between  $\phi(x_1), \phi(x_4)$ and $\phi(x_2), \phi(x_3)$. 
		The crossing equation in terms of $(u,v)$ and $(z, \bar{z})$ respectively are as follows,
		\be \label{Gzzb1}  \tilde{\mathcal G}(u,v) =\left(\frac{u}{v}\right)^{\Delta_\phi} \tilde{\mathcal G}(v,u), \quad\quad \tilde{\mathcal G}(z,\bar{z}) = \left( \frac{z\bar{z}}{(1-z)(1-\bar{z})} \right)^{\Delta_\phi} \tilde{\mathcal G}(1-z,1-\bar{z}).\ee
		
		
		\subsection{Analytic Functionals}
		
		One ingredient in our study is the analytic functionals introduced in \cite{mazavc2021basis}, which we will briefly mention here  referring to \cite{mazavc2021basis} for a more complete and rigorous description. It was conjectured that the four point correlation function of scalar field in a unitary theory can be expanded in terms of double-trace conformal blocks. In other words a correlation function ${\mathcal G} (z,\bar{z})$  in a unitary theory can be expanded in terms of the following basis set of functions,
		\be \{ G_{\Delta_{n,\ell}, \ell }^{(s)}(z,\bar{z}), \partial_\Delta G_{\Delta_{n,\ell}, \ell }^{(s)}(z,\bar{z}), G_{\Delta_{n,\ell}, \ell }^{(t)}(z,\bar{z}), \partial_\Delta G_{\Delta_{n,\ell}, \ell }^{(t)}(z,\bar{z}) \}. \ee 
		
		It stems from the study of correlation functions in CFTs in $d=1$ \cite{ mazac2016analytic, mazavc2019analytic, mazac2018analytic}. In the light of this study, they introduces ${\mathcal V}$, which denotes space of holomorphic functions ${\mathcal R}\times {\mathcal R} \rightarrow {\mathcal C}$, symmetric in two variables and bounded by a constant  away from $(z, \bar{z}) = (0,1)$. More precisely, ${\mathcal G}(z,\bar{z})\in {\mathcal V}$, if ${\mathcal G}(z,\bar{z})={\mathcal G}(\bar{z}, z)$ and $\forall \epsilon > 0$, $\exists$ $A >0$ such that $\forall$ $(z,\bar{z})$ satisfying $ |z|, |\bar{z}|, |1-z|, |1-\bar{z}| > \epsilon$,  we have $|{\mathcal G}(z,\bar{z}| < A$. One can further refine the functional space to be ${\mathcal U} \subset {\mathcal V}$, such that ${\mathcal G}(\bar{z}, z) \in {\mathcal U}$ if and only if  ${\mathcal G}(\bar{z}, z) \in {\mathcal V}$ and $\exists$ some constants $R>0$, $\epsilon>0$ and $A>0$ such that $\forall$ $(z,\bar{z})$ satisfying $|z|>R$, $|\bar{z}| > R$ we have ${\mathcal G}(\bar{z}, z) \leq A |z|^{-\frac{1}{2} - \epsilon}|\bar{z}|^{-\frac{1}{2} - \epsilon}$.
		
		In analogy with the behaviour in $d=1$, it is conjectured that  ${\mathcal G}(z,\bar{z})\in {\mathcal U}$ can be uniquely decomposed  as
		\be {\mathcal G}(z,\bar{z}) = {\mathcal G}^s (z,\bar{z}) + {\mathcal G}^t (z,\bar{z}),\ee
		where,  ${\mathcal G}^s (z,\bar{z}) , {\mathcal G}^t (z,\bar{z}) \in {\mathcal U}$ are Euclidean single valued around $(z, \bar{z})=(0,0)$ and $(1,1)$ respectively 
		with double discontinuity $dDisc_s\left[ {\mathcal G}^t (z,\bar{z})\right] =0$ and $dDisc_t \left[{\mathcal G}^s (z,\bar{z})\right] =0$. 

		Furthermore, ${\mathcal G}^t (z,\bar{z})$ can be expanded in terms of the s-channel double trace blocks and their $\Delta$-derivatives and similarly for ${\mathcal G}^s(z,\bar{z})$.
		The double trace blocks, organized by a pair of non-negative  integers, namely $n$ and spin $\ell$ correspond to the double trace operators  with scaling dimensions $\Delta_{n,\ell}= 2 \Delta_\phi + 2 n +\ell$. The s-channel double-trace blocks are holomorphic at $(z,\bar{z})=(0,0)$ and they constitute a basis for symmetric functions holomorphic in the neighbourhood of $(z,\bar{z})= (0,0)$ Similarly the t-channel double trace blocks  constitute a basis for symmetric functions holomorphic in the neighbourhood of $(z,\bar{z})= (1,1)$. In other words,
		\be {\mathcal G}^t (z,\bar{z})  = \sum\limits_{n,\ell}  a^s_{n,\ell} G_{\Delta_{n,\ell}, \ell }^{(s)}(z,\bar{z}) + b^s_{n,\ell} \partial_\Delta G_{\Delta_{n,\ell}, \ell }^{(s)}(z,\bar{z}), \quad 
		{\mathcal G}^s (z,\bar{z})  = \sum\limits_{n,\ell}  a^t_{n,\ell} G_{\Delta_{n,\ell}, \ell }^{(t)}(z,\bar{z}) + b^t_{n,\ell} \partial_\Delta G_{\Delta_{n,\ell}, \ell }^{(t)}(z,\bar{z}),\ee 
		
		Assembling these together it has been proposed that 
		\be \{ G_{\Delta_{n,\ell}, \ell }^{(s)}(z,\bar{z}), \partial_\Delta G_{\Delta_{n,\ell}, \ell }^{(s)}(z,\bar{z}), G_{\Delta_{n,\ell}, \ell }^{(t)}(z,\bar{z}), \partial_\Delta G_{\Delta_{n,\ell}, \ell }^{(t)}(z,\bar{z}) \}, \ee 
		constitute a basis for holomorphic functions in ${\mathcal U}$. It has been proved that the proposed basis is linearly independent.
		
		Considering the aforementioned basis set, one can introduce a set of functionals which are dual to the s-channel and t-channel double trace conformal block and their derivatives  \cite{mazavc2021basis}. This set of functionals, in turn constitute a  dual basis of the functionals. This basis of functionals dual to the conformal blocks in one-dimensional CFT was formulated and discussed in \cite{mazavc2019analytic,mazac2018analytic}. Later, it was developed in general dimension and is discussed in \cite{mazavc2021basis,paulos2020analytic}. 
		
		We denote the functional basis by $\{\alpha^{(s)}_{n,\ell},\alpha^{(t)}_{n,\ell},\beta^{(s)}_{n,\ell}, \beta^{(t)}_{n,\ell}\}$, as introduced in \cite{mazavc2021basis}. 
		The s-channel functionals act on the conformal blocks and satisfy the following duality properties,
		\begin{equation}\label{functional-duality-property1}
			\begin{split}
				\alpha^{(s)}_{n,\ell}\left[G_{\Delta_{n',\ell'},\ell'}^{(s)}\right]&= \delta_{nn'}\delta_{\ell,\ell'} ,\quad
				\alpha^{(s)}_{n,\ell}\left[\partial_\Delta G_{\Delta_{n',\ell'},\ell'}^{(s)}\right]= 0 ,\quad
				\alpha^{(s)}_{n,\ell}\left[G_{\Delta_{n',\ell'},\ell'}^{(t)}\right]= 0 ,\quad
				\alpha^{(s)}_{n,\ell}\left[\partial_\Delta G_{\Delta_{n',\ell'},\ell'}^{(t)}\right]= 0 ,\\
				\beta^{(s)}_{n,\ell}\left[G_{\Delta_{n',\ell'},\ell'}^{(s)}\right]&= 0,\quad
				\beta^{(s)}_{n,\ell}\left[\partial_\Delta G_{\Delta_{n',\ell'},\ell'}^{(s)}\right]= \delta_{nn'}\delta_{\ell,\ell'}, \quad
				\beta^{(s)}_{n,\ell}\left[G_{\Delta_{n',\ell'},\ell'}^{(t)}\right]= 0,\quad
				\beta^{(s)}_{n,\ell}\left[\partial_\Delta G_{\Delta_{n',\ell'},\ell'}^{(t)}\right]=0.
			\end{split}
		\end{equation}
		A similar set of t-channel functionals can also be introduced by interchanging
		$s \leftrightarrow t$. 
		
		These duality conditions imply that the action of s-channel functionals $ \alpha^{(s)}_{n,\ell}$ and $ \beta^{(s)}_{n,\ell}$ on t-channel conformal block $G^t_{\Delta, J}(z,\bar{z})$ has double zeroes on all double-trace spectrum and their action 
		on s-channel conformal blocks $G^s_{\Delta, J}(z,\bar{z})$ has double zeroes on all but one double-trace spectrum. The primal basis can be proved to be linearly independent and an explicit construction of these linear functionals is discussed in \cite{mazavc2021basis}.
		
		
		One more ingredient that we require to consider is the interpretation of the functional action  as the coefficients of conformal block decomposition in the exchange Witten diagram \cite{Zhou_2019}.  
		Exchange Witten  diagrams involve the propagation of intermediate states in the AdS/CFT framework. This computation can be related to the Polyakov blocks, which capture the complete contribution of intermediate states in conformal field theory\cite{mazavc2021basis}. In particular, we will briefly mention how functional actions on scalar conformal blocks can be written in terms of coefficients of exchange Witten  diagrams. 
		This has been discussed elaborately in \cite{mazavc2021basis}.
				
		If we restrict to s-channel conformal blocks $G^s_{\Delta, J}(z,\bar{z})$, with $\Delta_\phi > d/4$ so that $G^s_{\Delta, J}(z,\bar{z}) \in {\mathcal U}$, it is possible to write,
		\be\begin{split}\label{sch-conformal-block}
			G^s_{\Delta, J}(z,\bar{z}) &= \sum\limits_{n,\ell} \{ \alpha^{(s)}_{n,\ell}[G^s_{\Delta, J}(z,\bar{z})] G^s_{\Delta_{n,\ell}, \ell}(z,\bar{z})+ \beta^{(s)}_{n,\ell}[G^s_{\Delta, J}(z,\bar{z})] \partial_\Delta G^s_{\Delta_{n,\ell}, \ell}(z,\bar{z})\}
			\\& +
			\sum\limits_{n,\ell} \{ \alpha^{(t)}_{n,\ell}[G^s_{\Delta, J}(z,\bar{z})] G^t_{\Delta_{n,\ell}, \ell}(z,\bar{z})+ \beta^{(t)}_{n,\ell}[G^s_{\Delta, J}(z,\bar{z})] \partial_\Delta G^t_{\Delta_{n,\ell}, \ell}(z,\bar{z})\}
		\end{split} \ee
		where the expressions on the right hand side on both lines $\in {\mathcal U}$.
		
		By moving all the s-channel blocks on the left hand side and the t-channel blocks on the right hand side one can define s-channel Polyakov-Regge block as follows,
		\be
		P^s_{\Delta,J}(z,\bar{z}) =  G^s_{\Delta, J}(z,\bar{z}) - \sum\limits_{n,\ell} \{ \alpha^{(s)}_{n,\ell}[G^s_{\Delta, J}(z,\bar{z})] G^s_{\Delta_{n,\ell}, \ell}(z,\bar{z})+ \beta^{(s)}_{n,\ell}[G^s_{\Delta, J}(z,\bar{z})] \partial_\Delta G^s_{\Delta_{n,\ell}, \ell}(z,\bar{z})\},
		\ee
		and (\ref{sch-conformal-block}) implies that the s-channel Polyakov-Regge block admits the following t-channel expansion,
		\be
		P^s_{\Delta,J}(z,\bar{z}) =   \sum\limits_{n,\ell} \{ \alpha^{(t)}_{n,\ell}[G^s_{\Delta, J}(z,\bar{z})] G^t_{\Delta_{n,\ell}, \ell}(z,\bar{z})+ \beta^{(t)}_{n,\ell}[G^s_{\Delta, J}(z,\bar{z})] \partial_\Delta G^t_{\Delta_{n,\ell}, \ell}(z,\bar{z})\}
		\ee
		Establishing that $P^s_{\Delta,J}(z,\bar{z}) \in {\mathcal U}$ and that it admits the above s and t channel OPE amounts to the statement that $G_{\Delta, J}$ admits the expansion in (\ref{sch-conformal-block}). 
		
		This OPE structure is characteristic of exchange Witten diagram.  Consider the tree-level s-channel  exchange Witten diagram ${\mathcal W}^s_{\Delta,J}$ of a bulk field of dimension $\Delta$ and spin $J$ in AdS$_{d+1}$. The s-channel OPE of ${\mathcal W}^s_{\Delta,J}$ contains the single trace conformal blocks $G^s_{\Delta, J}(z,\bar{z})$, dressed by an infinite set of double-trace conformal blocks $G^s_{\Delta_{n,\ell}, \ell}(z,\bar{z})$ and $\partial_\Delta G^s_{\Delta_{n,\ell}, \ell}(z,\bar{z})$ with spins $\ell \leq J$. The t-channel OPE of ${\mathcal W}^s_{\Delta,J}$ contains only double-trace conformal blocks $G^t_{\Delta_{n,\ell}, \ell}(z,\bar{z})$ and $\partial_\Delta G^t_{\Delta_{n,\ell}, \ell}(z,\bar{z})$. One can normalize ${\mathcal W}^s_{\Delta,J}$ so that $G^s_{\Delta_{n,\ell}, \ell}(z,\bar{z})$ appears in the s-channel with unit coefficient. 
		The s and t channel OPE of s-channel Polyakov-Regge block has the correct  exchange Witten diagram as
		\be P^s_{\Delta,J}(z,{\bar z}) = A^{-1} {\mathcal W}^s_{\Delta,J}(z,{\bar z}) + {\mathcal C}(z,{\bar z}) \ee
		where $A$ is an overall normalization and ${\mathcal C}(z,{\bar z})$ is a finite linear combination of four-point contact diagrams. 
		By adding contact diagram one makes sure that the s-channel Polyakov-Regge block is inside ${\mathcal U}$.

		As explained in  \cite{mazavc2021basis} the scalar exchange Witten diagram admits the following conformal block decomposition
		\be\begin{split}
			{\mathcal W}^s_{\Delta,J=0} &= A G^s_{\Delta,0} + \sum\limits_{n=0}^\infty A_{n,0} G^s_{\Delta_{n,0}, 0 } +  \sum\limits_{n=0}^\infty D_{n,0} \partial G^s_{\Delta_{n,0}, 0 } 
			=  \sum\limits_{\ell=0}^\infty\sum\limits_{n=0}^\infty B_{n,\ell} G^t_{\Delta_{n,\ell}, \ell } +  \sum\limits_{\ell=0}^\infty\sum\limits_{n=0}^\infty C_{n,\ell} \partial G^t_{\Delta_{n,\ell}, \ell },\\
			{\mathcal W}^t_{\Delta,J=0} &= A G^t_{\Delta,0} + \sum\limits_{n=0}^\infty A_{n,0} G^t_{\Delta_{n,0}, 0 } +  \sum\limits_{n=0}^\infty D_{n,0} \partial G^t_{\Delta_{n,0}, 0 } 
			=  \sum\limits_{\ell=0}^\infty\sum\limits_{n=0}^\infty B_{n,\ell} G^s_{\Delta_{n,\ell}, \ell } +  \sum\limits_{\ell=0}^\infty\sum\limits_{n=0}^\infty C_{n,\ell} \partial G^s_{\Delta_{n,\ell}, \ell }.
		\end{split}\ee
		The explicit expressions of the coefficients as given in \cite{Zhou_2019}
		\be\begin{split} \label{coefficient}
			A &=  \frac{\pi^{d/2} \Gamma\left(\frac{\Delta_E}{2}\right)^4  \Gamma\left(\Delta_\phi - \frac{\Delta_E}{2}\right)^2  \Gamma\left(\Delta_\phi + \frac{\Delta_E - d }{2}\right)^2 }
			{8 \Gamma\left(\Delta_E \right) \Gamma\left(\Delta_\phi \right)^4\Gamma\left(\Delta_E + 1 - \frac{d}{2} \right)}\\
			A_{n,0} & =\frac{\pi^{d/2} \Gamma\left(n+\Delta_\phi\right)^4  \Gamma\left(-\frac{d}{2} + n + 2 \Delta_\phi\right)^2}
			{(n!)^2  \Gamma\left( \Delta_\phi \right)^4 \left(\Delta_E - 2 \Delta_\phi - 2 n \right)^2 \Gamma\left( 2(n+\Delta_\phi) \right) \left(-d  + \Delta_E + 2 \Delta_\phi + 2 n \right)^2 \Gamma\left( - \frac{d}{2} + 2 \Delta_\phi + 2 n \right) } \\
			& \times 
			\left(                          
			d - 4(\Delta_\phi + n) + (-\Delta_E + 2 \Delta_\phi + 2 n)(d - \Delta_E - 2 \Delta_\phi - 2 n) \right.  \\
			& \left.	 \left[        - \psi (-\frac{d}{2} + n+2 \Delta_\phi)  + \psi \left(      2 (n+\Delta_\phi) - \frac{d}{2}
			\right) 
			- 2 \psi (n+ \Delta_\phi) +  \psi (2(n+ \Delta_\phi)) + \psi (n+1)
			\right]
			\right)\\
			D_{n,0} &= \frac{
				\pi^{d/2} \Gamma\left(n+\Delta_\phi\right)^4  \Gamma\left(-\frac{d}{2} + n + 2 \Delta_\phi\right)^2 
			}
			{   
				(n!)^2  \Gamma\left( \Delta_\phi \right)^4 \left(- \Delta_E + 2 \Delta_\phi + 2 n \right) \Gamma\left( 2(n+\Delta_\phi)\right) (-d + \Delta_E  + 2 \Delta_\phi + 2 n) \Gamma\left( 2(n+\Delta_\phi) - \frac{d}{2} \right)
			},
		\end{split}
		\ee
		where $\psi(x)=\frac{\Gamma{}^'(x)}{\Gamma(x)}$ is the digamma function.		
		
		One can compare it with the $J=0$ Polyakov-Regge blocks
		\be 
		P^s_{\Delta,J=0} = \frac{1}{A} {\mathcal W}^s_{\Delta,J=0},\quad\quad P^t_{\Delta,J=0} = \frac{1}{A} {\mathcal W}^t_{\Delta,J=0},
		\ee
		which gives the following relations for the s-channel conformal block  $G^{(s)}_{\Delta_E,0}$
		\be\begin{split}\label{Witten-diagram}
			\alpha_{n,\ell}^{(s)}\left[G^{(s)}_{\Delta_E,0}\right] & = - \frac{A_{n,0}}{A}\delta_{l,0}, \qquad  \beta_{n,\ell}^{(s)} \left[G^{(s)}_{\Delta_E,0}\right] = -\frac{D_{n,0}}{A}\delta_{\ell,0}\\
			\alpha_{n,\ell}^{(t)}\left[G^{(s)}_{\Delta_E,0}\right] & =\frac{B_{n,\ell}}{A},  \qquad\qquad \beta_{n,\ell}^{(t)} \left[G^{(s)}_{\Delta_E,0}\right]  = \frac{C_{n,\ell}}{A}
		\end{split}\ee
		and similar  relations for the t-channel conformal block  $G^{(t)}_{\Delta_E,0}$ can be obtained by $s \leftrightarrow t$.
		
		In the first two equations of (\ref{Witten-diagram}), the parameters appearing on the right hand side are the coefficients of conformal blocks in the scalar exchange Witten diagram  in the direct channel. The explicit expressions of them are given in (\ref{coefficient}). On the other hand, there is no explicit expression for the parameters appearing on the right hand side of last two equations of (\ref{Witten-diagram}) as such. However, $B_{n,\ell}$ and $C_{n,\ell}$ follow some recursion relations as spelt out  in the appendix. We will use these relations to obtain the OPE coefficients and a detailed discussion of our method will be done in the next section.

		\section{Calculation of OPE coefficients}
	
		In this section, we apply the functionals to the mean field theory. As we will see, the application can determine the coefficients of operator product expansions, namely $\lambda_{n,l}$. This approach was used in the context of conformal field theory on ${\mathbb{R}\mathbf{P}}^d$ \cite{Giombi:2020xah} to determine the coefficients of operator product expansion. Our analysis will be quite similar and afterwards we will extend its application to the Wilson-Fisher model.
		
		\subsection{Mean field theory}
		
		We will consider mean field theory, one of the simplest examples of the conformal field theory \cite{Fitzpatrick_2013,Fitzpatrick:2011dm,Bissi_2020,Bissi:2022mrs,paulos2020analytic}. It is dual to the free field theories in the AdS, where all correlators are determined by the 2-point function of primary operators, the operators dual to the fields in the AdS \cite{Fitzpatrick:2011dm}. For example, four-point function of identical scalars with scaling dimension $\Delta_\phi$, involves in addition to the identity operator, a tower of intermediate operators $\phi\partial^{2n} \partial_{\mu_1} ... \partial_{\mu_\ell}\phi$. The latter are the double trace operators, (which we have already mentioned) characterized by a pair of integers, $n$ and $\ell$ and scaling dimension $\triangle_{n,\ell}= 2 \triangle_\phi + 2 n + \ell$. The engineering dimension is same as the conformal dimension of these operators. The correlation functions of these operators can be obtained by simply considering the Wick contractions.  Once we turn on interaction, the scaling dimension and the OPE coefficients will get modified. 
		
		In the mean field theory, the four point correlator of four identical scalars $\phi$ of dimension $\Delta_\phi$ can be expressed \cite{Gliozzi_2017,Bissi_2020,Bissi:2022mrs} 
		in terms of $(u,v)$ and $(z,\bar{z})$ respectively as (\ref{4point})
		where the function $\tilde{\mathcal G}$ is given by
		\be \label{Gzzb} \tilde{\mathcal G}(u,v) = 1 + u^{\Delta_\phi} + \left(\frac{u}{v}\right)^{\Delta_\phi}, \quad\quad  \tilde{\mathcal G} (z,\bar{z}) = 1 + (z\bar{z})^{\Delta_\phi} + \left(\frac{z\bar{z}}{(1-z)(1-\bar{z})}\right)^{\Delta_\phi} . \ee
		
		The modified conformal partial wave has become, as seen from (\ref{Gzzb})
		\be {\mathcal G}(z,\bar{z}) = 1 + (z\bar{z})^{-\Delta_\phi} +[ (1-z)(1-\bar{z})]^{-\Delta_\phi}  . \ee
		We would like to expand the CPW in terms of the conformal blocks. Comparing to (\ref{conformalblock}) and taking the redefinition of the conformal blocks into account, the second term can be easily identified as $G_{0,0}(z,\bar{z})$. 
		For the third term, we can make use of the fact that $(z\bar{z})^{\Delta_\phi} $ can be expanded in terms of the conformal blocks associated with the double trace operators as \cite{Gliozzi_2017},
		\be \label{Gliozziexpand}
		(z\bar{z})^{\Delta_\phi}=\sum_{n,\ell}^\infty \lambda_{n,\ell}g_{\Delta_{n,\ell}, \ell }^{(s)}(z,\bar{z}),
		\ee
		where we have introduced a superscript $s$ in the conformal blocks as they corresponds to the s-channel. $ \lambda_{n,\ell}$ are certain constant coefficients and their explicit expressions are given in \cite{Fitzpatrick:2011dm,Bissi_2020}. 
		\be\label{Kaplan's expression}
		\lambda_{n,\ell}=\frac{(-1)^\ell \left(\Delta_\phi-\frac{d}{2}+1\right)_n^2(\Delta_\phi)_{\ell+n}}{\ell ! n ! \left( \ell+\frac{d}{2}\right)_n\left(2\Delta_\phi+n-d+1\right)_n\left(2\Delta_\phi+2n+\ell-1\right)_\ell \left(2\Delta_\phi+n+\ell-\frac{d}{2}\right)_n}
		\ee 
		It has been shown in \cite{Dolan_2001} that the conformal blocks transform in the following manner,
		\be g^{(s)}_{\Delta,\ell}(u,v) = (-1)^\ell g^{(s)}_{\Delta,\ell}(\frac{u}{v}, \frac{1}{v}) \ee
		Using this relation we obtain from (\ref{Gliozziexpand})
		\be 
		[(1-z)(1- \bar{z})]^{- \Delta_\phi}=\sum_{n,\ell}^\infty \lambda_{n,\ell} (-1)^\ell G_{\Delta_{n,\ell}, \ell }^{(s)}(z,\bar{z}). \ee
		where the redefined conformal block refers to the s-channel only. 
		Therefore we can write
		\be \label{4-point} {\mathcal G}(z,\bar{z}) = 1 + G_{0,0}^{(s)}(z,\bar{z}) +  \sum_{n,\ell}^\infty (-1)^\ell \lambda_{n,\ell}G_{\Delta_{n,\ell}, \ell }^{(s)}(z,\bar{z}).\ee
		
		The above expression can also be written in terms of the t-channel conformal blocks. The t-channel and s-channel conformal blocks are related to each other through the following transformation,
		\be G_{\Delta_{n,\ell}, \ell }^{(s)}(z,\bar{z}) = G_{\Delta_{n,\ell}, \ell }^{(t)}(1-z,1-\bar{z}).\ee 
		Using the crossing symmetry of the four point function, and (\ref{4-point}) we can write down the same expression in terms of s-channel and t-channel leading to the following equality
		\be \label{unperturbed-crossing} G_{0,0}^{(s)}(z,\bar{z}) +  \sum_{n,\ell}^\infty (-1)^\ell \lambda_{n,\ell} G_{\Delta_{n,\ell}, \ell }^{(s)}(z,\bar{z}) =  G_{0,0}^{(t)}(z,\bar{z}) +  \sum_{n,\ell}^\infty (-1)^\ell \lambda_{n,\ell}G_{\Delta_{n,\ell}, \ell }^{(t)}(z,\bar{z}).\ee
		This expression can be considered as crossing symmetry relation.
		

		To begin with we consider crossing symmetry of the four-point correlation function expanded in terms of the conformal blocks associated with double trace operators, {\it i.e.} (\ref{unperturbed-crossing}). Applying the $\alpha^{s}_{n,\ell}$-functionals on both sides of the equation (\ref{unperturbed-crossing}),
		\be \label{alpha-crossing} \alpha_{n,\ell}^s \left(G_{0,0}^{(s)}(z,\bar{z})\right) +  \sum_{m,k}^\infty (-1)^k \lambda_{m,k}  \alpha_{n,\ell}^s\left( G_{\Delta_{m,k}, k}^{(s)}(z,\bar{z}) \right)=   \alpha_{n,\ell}^s \left(G_{0,0}^{(t)}(z,\bar{z})\right) +  \sum_{m,k}^\infty (-1)^k \lambda_{m,k} \alpha_{n,\ell}^s\left(G_{\Delta_{m,k}, k }^{(t)}(z,\bar{z})\right).\ee
		
		Using (\ref{Witten-diagram}) and (\ref{functional-duality-property1})  to reduce the first and the second term respectively of both sides of the above equation we obtain, 
		\begin{equation}\label{alpha-functional-sum-rule1}
			\left[-\frac{A_{n,0}}{A}\right]_{\Delta_E=0}\delta_{\ell,0}+(-1)^\ell \lambda_{n,\ell} = \left[\frac{B_{n,\ell}}{A}\right]_{\Delta_E=0}.
		\end{equation}
		From the explicit expression (\ref{coefficient}), it is seen that at the limit $\Delta_E\rightarrow 0$, $A$ diverges, while $A_{n,0}$ remains finite. Thus, the first term  of (\ref{alpha-functional-sum-rule1}) vanishes and the equation becomes:
		\begin{equation}\label{alpha-functional-sum-rule2}
			(-1)^\ell \lambda_{n,\ell} = \left[\frac{B_{n,\ell}}{A}\right]_{\Delta_E=0}
		\end{equation}
		We observe that, \ref{alpha-functional-sum-rule2} implies that the OPE coefficients obeys the same recursion relation as that of $b_{n,\ell}$ (\ref{recursion-relation-bnl}). 
		
		Next we apply the $\beta_{n,l}^s$-functionals on both sides of the equation (\ref{unperturbed-crossing}),
		\be \label{beta-crossing} \beta_{n,\ell}^s(G_{0,0}^{(s)}(z,\bar{z})) +  \sum_{m,k}^\infty (-1)^k \lambda_{m,k}  \beta_{n,\ell}^s(G_{\Delta_{m,k}, k}^{(s)}(z,\bar{z})) =   \beta_{n,\ell}^s(G_{0,0}^{(t)}(z,\bar{z})) +  \sum_{m,k}^\infty (-1)^k \lambda_{m,k} \beta_{n,\ell}^s(G_{\Delta_{m,k}, k }^{(t)}(z,\bar{z})).\ee
		Using (\ref{Witten-diagram}) we obtain
		\begin{equation}\label{alpha-functional-sum-rule3}
			\left[-\frac{D_{n,0}}{A}\right]_{\Delta_E=0}\delta_{\ell,0}=\left[\frac{C_{n,\ell}}{A}\right]_{\Delta_E=0}
		\end{equation}
		
		Once again, referring to (\ref{coefficient}), we observe that at the limit $\Delta_E\rightarrow 0$, $A$ diverges, while $D_{n,0}$ remains finite. Hence, left hand side  of above equation \ref{alpha-functional-sum-rule3} becomes zero and we obtain:
		\begin{equation}\label{alpha-functional-sum-rule4}
			\left[\frac{C_{n,\ell}}{A}\right]_{\Delta_E=0}=0
		\end{equation}
		
		In order to determine $\lambda_{n,\ell}$, we consider the recursion relation obeyed by $b_{n,\ell}$, given in (\ref{recursion-relation-bnl}). 
		Since $\lambda_{n,\ell}$ is the coefficient arising in the expansion of conformal block in terms of the double trace operator in the context of mean field theory, we assume that for non-zero values of $n$, $\lambda_{n,\ell}$ will vanish. This is in the spirit of the assumption that nonzero contribution to $\lambda_{n,\ell}$ for $n\neq 0$ begins  from the order $0(\epsilon^2)$ \cite{Bissi:2022mrs}. They have considered nonzero contribution to the conformal partial wave (for unperturbed mean field theory) from the identity operator and intermediate operators having twist $\tau = d-2$, which corresponds to $n=0$,  With this assumption, the recursion relation reduces to
		
		\begin{equation} \label{recursion-relation-b0l}
			- 2 (\ell+2)^2 b_{0 , \ell+1} - \frac{ (\ell-1)^2 \ell^2 } {2 (2 \ell -1)(2\ell+1) } b_{0,\ell-1} + (4+\ell+\ell^2) b_{0,\ell} = 0.
		\end{equation}
		One can show that the equation can be solved to obtain 
		\be b_{0,\ell} = \frac{\Gamma(\ell+1)^2}{\Gamma(2 \ell+1)}\ee
		This results match with (\ref{Kaplan's expression}) \cite{Fitzpatrick:2011dm,Bissi_2020} for $\Delta_\phi=\frac{d}{2}-1$, $d=4$.. 

		\subsection{Wilson-Fisher model}
		
		In this subsection we will consider perturbations around the mean field theory and determine the correction in the coefficient of operator product expansion. In particular we will use the Wilson-Fisher fixed point \cite{Giombi:2020xah,Rychkov:2015naa, Basu:2015gpa, Raju:2015fza, Gliozzi:2016ysv, Dey:2016mcs, Roumpedakis:2016qcg, Alday:2016jfr, Gopakumar:2016wkt, Gopakumar:2016cpb, Gliozzi:2017hni, Liendo:2017wsn, Dey:2017fab, Gliozzi:2017gzh,Alday:2017zzv, Gopakumar:2018xqi}, which may be defined in $d=4-\epsilon$ dimension. The Lagrangian can be written as
		\be
		S= \frac{\Gamma\left(\frac{d}{2}-1\right)}{4\pi^{\frac{d}{2}}}  \int d^dx \left( \frac{1}{2} (\partial \phi^I)^2 + \frac{g}{4} (\phi^I\phi^I)   \right)  \ee
		which describes perturbation of a theory of N free fields.
		
		The perturbation induces corrections to the scaling dimension of the external field $\phi$, which is given as
		\begin{equation}\label{Delta-phi-WF}
			\Delta_\phi=\frac{d}{2}-1+\frac{\epsilon^2}{108}+O(\epsilon^3) =  \Delta_\phi^{(0)}+	\frac{\epsilon^2}{108}+O(\epsilon^3),
		\end{equation}
		where we have introduced $\Delta_\phi^{(0)} = \frac{d}{2}-1$
		
		The small parameter $\epsilon$ plays the role of perturbation and for $\epsilon$ equals zero, the theory gets back to the unperturbed mean field theory \cite{rychkov2015epsilonexpansionconformalfieldtheory}. 
		Assuming the perturbation is small, the OPE coefficients associated with the conformal blocks and the scaling dimension of the exchange operator can be expanded in a power series in $\epsilon$  around the mean field theory, as follows:
		\begin{align}
			\lambda_{n,\ell}&=\lambda_{n,\ell}^{(0)}+\epsilon \lambda_{n,\ell}^{(1)}+O(\epsilon^2)\label{OPE correction}\\
			\Delta_{n,\ell}&=\Delta_{n,\ell}^{(0)}+\epsilon \gamma_{n,\ell}^{(1)}+O(\epsilon^2)\label{anomalous dimension}
		\end{align}
		
		As we did in the unperturbed case, we will consider the crossing symmetry equation (\ref{unperturbed-crossing}) and consider corrections of various terms in the crossing symmetry equation upto first order and obtain
		\begin{equation}
			\begin{split}\label{perturbed-crossing}
				& \left(G_{0,0}^{(s)}\right)^{O(\epsilon^0)}+\left(G_{0,0}^{(s)}\right)^{O(\epsilon)} + \sum_{n,\ell} (-1)^\ell \left( \lambda_{n,\ell}^{(0)} + \epsilon \lambda_{n,\ell}^{(1)} \right) G_{\Delta_{n,\ell }+ \epsilon \gamma_{n,\ell}^{(1)},l }^{(s)} \\
				& = 
				\left(G_{0,0}^{(t)}\right)^{O(\epsilon^0)} + \left(G_{0,0}^{(t)}\right)^{O(\epsilon)} + \sum_{n,\ell} (-1)^\ell \left( \lambda_{n,\ell}^{(0)} + \epsilon \lambda_{n,\ell}^{(1)} \right) G_{\Delta_{n,\ell } +\epsilon \gamma_{n,\ell}^{(1)}, l }^{(t)} 
		\end{split}\end{equation}
		
		We will consider the action of $\alpha_{n,l}^s$-functional on the crossing symmetry equation of the perturbed case (\ref{perturbed-crossing}) and obtain
		\begin{equation}
			\begin{split}\label{alpha-functional-sum-rule-WF}
				& \alpha_{mk}^{(s)}\left( \left(G_{0,0}^{(s)}\right)^{O(\epsilon^0)}\right) 
				+ \alpha_{mk}^{(s)}\left( \left(G_{0,0}^{(s)}\right)^{O(\epsilon)}\right) + \sum_{n,\ell} (-1)^\ell \left( \lambda_{n,\ell}^{(0)} + \epsilon \lambda_{n,\ell}^{(1)} \right)  \alpha_{mk}^{(s)}\left(G_{\Delta_{n,\ell }+ \epsilon \gamma_{n,\ell}^{(1)}, \ell}^{(s)}\right) \\
				& = 
				\alpha_{mk}^{(s)}\left(\left(G_{0,0}^{(t)}\right)^{O(\epsilon^0)} \right) +  \alpha_{mk}^{(s)}\left(\left(G_{0,0}^{(t)}\right)^{O(\epsilon)} \right) + \sum_{n,\ell} (-1)^\ell \left( \lambda_{n,\ell}^{(0)} + \epsilon \lambda_{n,\ell}^{(1)} \right)  \alpha_{mk}^{(s)}\left(G_{\Delta_{n,\ell } +\epsilon \gamma_{n,\ell}^{(1)},\ell}^{(t)} \right)
		\end{split}\end{equation}
		
		We will be interested only in the $O(\epsilon)$ contributions on both sides. Furthermore, since we have assumed that for $n\neq 0$  $\lambda_{n,\ell}$ receive contribution only at the quadratic order of $\epsilon$ or higher we can impose $\lambda_{n,\ell}^{(0)} = \lambda_{0,l}^{(0)} \delta_{n,0}$ and $\lambda_{n,\ell}^{(1)} = \lambda_{0,l}^{(1)}\delta_{n,0}$. Substituting in the above expression, we obtain
		\begin{equation}
			\begin{split}\label{alpha-functional-sum-rule-WF}
				& \alpha_{mk}^{(s)}\left( \left(G_{0,0}^{(s)}\right)^{O(\epsilon)}\right) + \sum_\ell (-1)^\ell \left( \lambda_{0,\ell}^{(0)} + \epsilon \lambda_{0,\ell}^{(1)} \right)  \alpha_{mk}^{(s)}\left(G_{\Delta_{0,\ell }+ \epsilon \gamma_{0,\ell}^{(1)},\ell}^{(s)}\right) \\
				& = 
				\alpha_{mk}^{(s)}\left(\left(G_{0,0}^{(t)}\right)^{O(\epsilon)} \right) + \sum_\ell (-1)^\ell \left( \lambda_{0,\ell}^{(0)} + \epsilon \lambda_{0,\ell}^{(1)} \right)  \alpha_{mk}^{(s)}\left(G_{\Delta_{0,\ell } +\epsilon \gamma_{0,\ell}^{(1)},\ell}^{(t)} \right)
		\end{split}\end{equation}
		
		Segregating only contributions of $0(\epsilon)$,
		\begin{equation}
			\begin{split}\label{alpha-functional-sum-rule-WF}
				& \alpha_{mk}^{(s)}\left( \left(G_{0,0}^{(s)}\right)^{O(\epsilon)}\right) + \sum_\ell (-1)^\ell  \lambda_{0,\ell}^{(0)}  \alpha_{mk}^{(s)}\left(\left(G_{\Delta_{0,\ell }+ \epsilon \gamma_{0,\ell}^{(1)},\ell}^{(s)}\right)^{O(\epsilon)}\right)  + \sum_\ell (-1)^\ell  \lambda_{0,\ell}^{(1)}   \alpha_{mk}^{(s)}\left(G_{\Delta_{0,\ell },\ell}^{(s)}\right) \\
				& = 
				\alpha_{mk}^{(s)}\left(\left(G_{0,0}^{(t)}\right)^{O(\epsilon)} \right) +  \sum_\ell (-1)^\ell  \lambda_{0,\ell}^{(0)}  \alpha_{mk}^{(s)}\left(\left(G_{\Delta_{0,\ell }+ \epsilon \gamma_{0,\ell}^{(1)},\ell}^{(t)}\right)^{O(\epsilon)}\right)  + \sum_\ell (-1)^\ell  \lambda_{0,\ell}^{(1)}   \alpha_{mk}^{(s)}\left(G_{\Delta_{0,\ell },\ell}^{(t)}\right)
		\end{split}\end{equation}
		
		From duality $\alpha_{mk}^{(s)}\left(G_{\Delta_{0,\ell },\ell}^{(s)}\right)= \delta_{m,0}\delta_{\ell,k}$ and $\alpha_{mk}^{(s)}\left(G_{\Delta_{0,\ell },\ell}^{(t)}\right)= 0$, the above equation will further get simplified as
		\begin{equation}
			\begin{split}\label{alpha-functional-sum-rule-WF}
				& \alpha_{mk}^{(s)}\left( \left(G_{0,0}^{(s)}\right)^{O(\epsilon)}\right) + \sum_\ell (-1)^\ell  \lambda_{0,\ell}^{(0)}  \alpha_{mk}^{(s)}\left(G_{\Delta_{0,\ell }+ \epsilon \gamma_{0,\ell}^{(1)},\ell}^{(s)}\right)^{O(\epsilon)}  + (-1)^k  \lambda_{0,k}^{(1)}   \delta_{m,0} \\
				& = 
				\alpha_{mk}^{(s)}\left(\left(G_{0,0}^{(t)}\right)^{O(\epsilon)} \right) +  \sum_\ell (-1)^\ell  \lambda_{0,\ell}^{(0)}  \alpha_{mk}^{(s)}\left(G_{\Delta_{0,\ell }+ \epsilon \gamma_{0,\ell}^{(1)},\ell}^{(t)}\right)^{O(\epsilon)}  
		\end{split}\end{equation}
		
		We further assume $\gamma_{0,\ell}^{(1)} = \gamma_{0,0}^{(1)}\delta_{\ell,0}$ \cite{Alday_2017} and considering only the $O(\epsilon)$ part of the equation we get
		\begin{equation}
			\begin{split}\label{alpha-functional-sum-rule-WF}
				& \alpha_{mk}^{(s)}\left( \left(G_{0,0}^{(s)}\right)^{O(\epsilon)}\right) +  \lambda_{0,0}^{(0)}  \alpha_{mk}^{(s)}\left(G_{\Delta_{0,0}+ \epsilon \gamma_{0,0}^{(1)},0}^{(s)}\right)^{O(\epsilon)}  + (-1)^k  \lambda_{0,k}^{(1)}   \delta_{m,0} \\
				& = 
				\alpha_{mk}^{(s)}\left(\left(G_{0,0}^{(t)}\right)^{O(\epsilon)} \right) +   \lambda_{0,0}^{(0)}  \alpha_{mk}^{(s)}\left(G_{\Delta_{0,0 }+ \epsilon \gamma_{0,0}^{(1)},0}^{(t)}\right)^{O(\epsilon)}  
		\end{split}\end{equation}
		
		The action of $\alpha_{n,\ell}^s$ -functional on various conformal blocks as above will be determined by using (\ref{Witten-diagram}). 
		\begin{equation}
			\begin{split}  \label{alpha-functional-sum-rule-WF1}
				&	- \left(   \frac{A_{m,0}  }{A}  \right)^{O(\epsilon)}_{\Delta_E=0} \delta_{k,0}   - \left(   \frac{A_{m,0}  }{A}  \right)^{O(\epsilon)}_{\Delta_E=\Delta_{0,0}+ \epsilon \gamma_{0,0}^{(1)}} \delta_{k,0}
				+ \lambda_{m,k}^{(1)} (-1)^k  \\
				& = 
				\left(   \frac{B_{m,k}}{A}  \right)^{O(\epsilon)}_{\Delta_E=0} 
				+  \left(   \frac{B_{m,k}}{A}  \right)^{O(\epsilon)}_{\Delta_E=\Delta_{0,0}+ \epsilon \gamma_{0,0}^{(1)}} 
		\end{split}\end{equation}
		where we can determine $A_{m,0}$ and $A$ from (\ref{coefficient}) by setting $\Delta_E=0$ for the first term and ${\Delta_E=\Delta_{0,0}+ \epsilon \gamma_{0,0}^{(1)}}$ for the second term on left hand side. $\Delta_\phi$ is given in (\ref{Delta-phi-WF}). Since, for $\Delta_E=0$, $A$ diverges but $A_{m,0}$ remains finite, the first term on the left had side will vanish. From (\ref{alpha-functional-sum-rule2}) and (\ref{Kaplan's expression}) one can obtain $\left(\frac{B_{m,k}}{A}  \right)^{O(\epsilon)}_{\Delta_E=0}$.
		The expression is
		\be \label{alpha-functional-sum-rule-WF2}
		\left(\frac{B_{m,k}}{A}  \right)^{O(\epsilon)}_{\Delta_E=0} =\left(\frac{\Gamma(1+k)^2(-2h_k + h_{2k})}{\Gamma(1+2k)}\right)\delta_{m,0}.
		\ee
		where $h_k = \sum\limits_{n=1}^k \frac{1}{n}$ represents the harmonic numbers. Second term of left hand side is explicitly calculated and it is found that
		\be \label{alpha-functional-sum-rule-WF3}
		\left(-\frac{A_{m,0}}{A}\right)_{\Delta_E=2\Delta_\phi+\epsilon\gamma_{0,0}^{(1)}}^{O(\epsilon)}\delta_{k,0}=\left(\gamma_{0,0}^{(1)}\right)^2 \delta_{k,0}\delta_{m,0}.
		\ee
		Since we have assumed that non-zero contribution of $\lambda_{n,\ell}$  for $n\neq 0$ begins from quadratic order of $\epsilon$, $\lambda^{(1)}_{m,k} = \lambda^{(1)}_{0,k} \delta_{m,0}$. Since all the terms in (\ref{alpha-functional-sum-rule-WF2}), are proportional to $\delta_{m,0}$ except the last term, $\left(\frac{B_{m,k}}{A}  \right)^{O(\epsilon)}_{\Delta_E=\Delta_{0,0}+ \epsilon \gamma_{0,0}^{(1)}}$, this also should vanish unless $m=0$. As explained in the appendix from the recursion relation one can obtain 
		\be \label{alpha-functional-sum-rule-WF4}
		\left(\frac{B_{m,k}}{A}  \right)^{O(\epsilon)}_{\Delta_E=\Delta_{0,0}+ \epsilon \gamma_{0,0}^{(1)}} = \left(\frac{B_{0,0}}{A}\right)^{O(\epsilon)}\delta_{m,0}\delta_{k,0}.
		\ee
		Assembling all the parts together, we obtain
		\be\label{final result}
		(-1)^k \lambda^{(1)}_{m,k}  =\left(\gamma^{(1)}_{0,0}\right)^2 \delta_{m,0} \delta_{k,0}+ \frac{\Gamma(1+k)^2(-2h_k + h_{2k})}{\Gamma(1+2k)}\delta_{m,0}  +  \frac{B^{(1)}_{0,0}}{A}\delta_{m,0}\delta_{k,0}.
		\ee
		
		This expression can be compared with the values obtained in literature using other methods and one can find precise agreement\cite{Gopakumar:2016cpb,Bissi_2020}.
		
		Using (\ref{final result}), the correction in OPE coefficients at the order of $O(\epsilon)$ for all the conformal blocks, except for $n=0$ and $l=0$ can be determined. It has been found that for any value of $n$, the correction in OPE coefficients can be obtained by:
		\be
		(-1)^l \lambda_{n,l}^{(1)} =\left(\frac{\Gamma(1+l)^2(-2h_l + h_{2l})}{\Gamma(1+2l)}\right)\delta_{n,0} \quad \text{for} \quad l\ge 1
		\ee
		This expression is same as the OPE coefficients in mean field theory as given in (\ref{Kaplan's expression}) with $d=4-\epsilon$ and $\Delta_\phi=\frac{d}{2}-1=1-\frac{\epsilon}{2}$. 		
		
		\section{Conclusion}
		
		One of the approaches in the analytic bootstrap is to use double twist  functionals. A complete basis of such analytic functionals are constructed in general dimension. Applying these functionals on correlators of conformal field theories leads to strong constraints on CFT data. We have applied this analytic functionals  in the case of mean field theory and Wilson-Fisher model. Following  \cite{Zhou_2019} and comparing with the exchange Witten diagrams,  with a few additional assumptions,  we have obtained the OPE coefficients and corrections in the conformal dimension upto first order. A comparison with these expressions obtained through other methods shows precise agreement. This has demonstrated application of analytic functional approach \cite{mazavc2021basis} to specific models. 
		
		In our discussion we restricted ourselves only to the first order corrections in $\epsilon$. A natural extension of our work is to consider higher order corrections.  Using the recursion relations it is certainly possible to obtain the higher order corrections to the OPE coefficients, with minimal seed value of some initial coefficients. For the sake of simplicity we have considered all the operators with equal conformal dimensions. This can be generalized to the cases where the operators have different weights.
		
		In order to have an application to a more realistic context one needs to consider the theories which involves fermions, such as Gross-Naveu model or Quantum Chromodynamics. This approach can also be extended to the supersymmetric theories. In particular, study of ${\mathcal N}=4$ super Yang-Mills theory in the $1/N$ expansion \cite{Rastelli:2016nze, Alday:2017xua, Aprile:2017bgs, Aprile:2017xsp, Rastelli:2017udc, Aprile:2017qoy,Alday:2017vkk,Aprile:2018efk,Caron-Huot:2018kta,Alday:2018kkw,Alday:2018pdi,Binder:2019jwn,Goncalves:2019znr,Drummond:2019odu} may be interesting \cite{mazavc2021basis}. The functional approach may provide  more systematic method compared to the other approaches. 
		
		\section*{Acknowledgement}
	Part of this work was presented at the Annual Conference of the
	Physics Academy of North East (PANE), 2024 organised by University of Science and Technology, Meghalaya (USTM). BJK is thankful to the local organizing committee for the hospitality during the stay at USTM.		
		\appendix 
		\section{\bf Recursion Relations}
		
		We have found that the OPE coefficients in the expansion of the conformal blocks in terms of the double trace operators are related to the coefficient of the exchange Witten diagram. For some coefficients has an expressions in closed form, while for the other one can obtain a recursion relation in general.  Considering the action of t-channel equation of motion operator, these recursion relations have been obtained among the coefficient of the exchange Witten diagram and the details are given in  \cite{Zhou_2019}.  In this Appendix we will review those relations and will discuss how these recursion relations can be used to obtain closed form expression of the expansion coefficients.
		
		It was shown that action of t-channel equation of motion operator on the derivative block leads to relations, which involves the following parameter:
		\begin{equation}\begin{split} \label{expression-of RSTUW}
				\mathcal{R}&=- \frac{(\ell+2\epsilon')(-\Delta+2\Delta_\phi+l)^2}{2(l+\epsilon')}\\
				\mathcal{S}&=- \frac{\ell(\Delta-2\Delta_\phi+l+2\epsilon')^2}{2(l+\epsilon')}\\
				\mathcal{T}&=- \frac{(\Delta-1)(\Delta-2\epsilon')(\Delta+\ell)^2(l+2\epsilon')(\Delta+2\Delta_\phi+l-2\epsilon'-2)^2}{32(\Delta-\epsilon'-1)(\Delta-\epsilon')(\Delta+l-1)(\Delta+\ell+1)(l+\epsilon')}\\
				\mathcal{U}&=- \frac{(\Delta-1)\ell(\Delta-2\epsilon')(-\Delta+\ell+2\epsilon')^2(-\Delta-2\Delta_\phi+\ell+4\epsilon'+2)^2}{32(\Delta-\epsilon'-1)(\Delta-\epsilon')(l+\epsilon')(-\Delta+l+2\epsilon'-1)(-\Delta+l+2\epsilon'+1)}\\
				\mathcal{W}&=C^{(2)}_{\Delta_E,\ell_E}+\frac{1}{2}C^{(2)}_{\Delta,\ell}-2C^{(2)}_{\Delta_\phi,0}.\end{split}\end{equation}	
		Where, $C^{(2)}_{\Delta,l}=\Delta(\Delta-d)+l(l+d-2)$ are the eigenvalue of quadratic conformal Casimir operator and  $\epsilon'=\frac{d}{2}-1$.
		
		The recursion relations are expressed in terms of the parameters  $\{\mathcal{R}_{n,\ell}$, $\mathcal{S}_{n,\ell}$, $\mathcal{T}_{n,\ell}$, $\mathcal{U}_{n,\ell}$ and $ \mathcal{W}_{n,\ell}\}$, which can be obtained by substituting $\Delta=2\Delta_\phi+2n+l$ in (\ref{expression-of RSTUW}).  There is another set of parameters,  $\mathcal{R'}_{n,\ell}$ etc., which can be obtained by partial differentiating equations of (\ref{expression-of RSTUW}) with respect to $\Delta$ and evaluating them at $\Delta=2\Delta_\phi+2n+l$. With these parameters one can write down the relations as
		\begin{equation} \begin{split}
				\label{recursion-relation-bc-only}
				\mathcal{R}_{n+1,\ell-1}\mathcal{C}_{n+1,\ell-1}+\mathcal{S}_{n,\ell+1}\mathcal{C}_{n,\ell+1}+\mathcal{T}_{n,\ell-1}\mathcal{C}_{n,\ell-1}+\mathcal{U}_{n-1,\ell+1}\mathcal{C}_{n-1,\ell+1}+\mathcal{W}_{n,\ell}\mathcal{C}_{n,\ell} & =\tilde{c}_{n,\ell}	,
				\\
				\mathcal{R}_{n+1,\ell-1}\mathcal{B}_{n+1,\ell-1}+\mathcal{S}_{n,\ell+1}\mathcal{B}_{n,\ell+1}+\mathcal{T}_{n,\ell-1}\mathcal{B}_{n,\ell-1}+\mathcal{U}_{n-1,\ell+1}\mathcal{B}_{n-1,\ell+1}+ \mathcal{W}_{n,\ell}\mathcal{B}_{n,\ell} & \\
				+\mathcal{R'}_{n+1,\ell-1}\mathcal{C}_{n+1,\ell-1}+\mathcal{S'}_{n,\ell+1}\mathcal{C}_{n,\ell+1}+\mathcal{T'}_{n,\ell-1}\mathcal{C}_{n,\ell-1}+\mathcal{U'}_{n-1,\ell+1}\mathcal{C}_{n-1,\ell+1}+\mathcal{W'}_{n,\ell}\mathcal{C}_{n,\ell} &=\tilde{a}_{n,\ell}
			\end{split}
		\end{equation}
		In above equations (\ref{recursion-relation-bc-only}),  $\mathcal{B}_{n,\ell}$ and $\mathcal{C}_{n,\ell}$ are related to the exchange Witten diagram coefficients in crossed channel, $B_{n,\ell}$ and $C_{n,\ell}$ via:
		\begin{equation}\label{curlybc-straight b,c}
			\mathcal{B}_{n,\ell}\mathcal{N}_{\epsilon',\ell}=B_{n,\ell}, \qquad \mathcal{C}_{n,\ell}\mathcal{N}_{\epsilon',\ell}=C_{n,\ell}
		\end{equation}
		where $\mathcal{N}_{\epsilon',\ell}=\frac{(\epsilon')_\ell}{(2 \epsilon')_\ell}$. 
		
		In order to determine the recursion relation satisfied by  the coefficients of the expansion of the conformal blocks in terms of double trace operators,  $\lambda_{n,\ell}$  we consider (\ref{recursion-relation-bc-only}), divide both sides by $A$. 
		
		For our purpose, we can reduce it further by defining $b_{n,\ell} = \left(\frac{B_{n,\ell}}{A}\right)_{\Delta_E=0}$ and  $c_{n,\ell} = \left(\frac{C_{n,\ell}}{A}\right)_{\Delta_E=0}$. The recursion relation changes as we move from mean field theory to perturbed model. We will consider the unperturbed and perturbed models in turn.\\
		
		$\bullet$ {\bf Unperturbed model:}  From \crefrange{alpha-functional-sum-rule1}{alpha-functional-sum-rule4} we observe that we require recursion relations for $\Delta_E=0$ .
		
		By dividing (\ref{recursion-relation-bc-only}) with $A$ and setting $\Delta_E=0$, we obtain a set of recursion relations for $b_{n,\ell}$ and  $c_{n,\ell}$. Since $A$ diverges for  $\Delta_E=0$, we set  $\frac{\tilde{c}_{n,\ell}}{A}=0$ and $\frac{\tilde{a}_{n,\ell}}{A}=0$. From (\ref{alpha-functional-sum-rule4}), we obtain $c_{n,\ell}=0$ and this is consistent with the second equation of (\ref{recursion-relation-bc-only}). Substituting that in the first equation it reduces to 
		\begin{equation}\label{recursion-relation-b-mod}
			\mathcal{R}_{n+1,\ell-1} b_{n+1,\ell-1} + \mathcal{S}_{n , \ell+1} b_{n , \ell+1} + \mathcal{T}_{ n, \ell-1} b_{n , \ell-1} + \mathcal{U}_{n-1,\ell+1}b_{n-1,\ell+1} + \mathcal{W}_{n ,\ell} b_{n ,\ell} = 0
		\end{equation}
		This provides us a recursion relation for $b_{n,\ell}$, which as per (\ref{alpha-functional-sum-rule2}) can be identified with
		$(-1)^\ell \lambda_{n , \ell}$.
		
		Since for the unperturbed case, we have $d=4$, $\Delta_\phi=\frac{d}{2}-1=1$ and $\epsilon'=1$, $\mathcal{R}_{n,\ell}$, $\mathcal{S}_{n,\ell}$ etc., for $\Delta_E=0$ are given by
		\begin{equation}\begin{split}\label{R(n,l)-unperturbed}
				\mathcal{R}_{n,\ell}&=-\frac{2(2+\ell)n^2}{1+\ell} 
				\\
				\mathcal{S}_{n,\ell}&=-\frac{2(\ell(1+\ell+n)^2)}{1+\ell}\\
				\mathcal{T}_{n,\ell}&=-\frac{(2+\ell)(\ell+n)^2(1+\ell+n)^2}{2(1+\ell)(1+2\ell+2n)(3+2\ell+2n)}\\
				\mathcal{U}_{n,\ell}&=-\frac{\ell(-1+n)^2n^2}{2(1+\ell)(-1+2n)(1+2n)}\\
				\mathcal{W}_{n,\ell}&=4 + \ell + \ell^2 + 2 \ell n + 2 n^2
		\end{split}\end{equation}
		
		This leads to the following recursion relation
		\begin{equation}\begin{split} \label{recursion-relation-bnl}
				&2 (n+1)^2 b_{n+1,\ell-1} + 2 (n+\ell+2)^2 b_{n , \ell+1} + \frac{(n+\ell-1)^2 (n+\ell)^2}{2 ( 2n+2 \ell -1)(2n+2\ell+1)} b_{n,\ell-1}\\
				& + \frac{(n-2)^2(n-1)^2}{2(2n-3)(2n-1)} b_{n-1,\ell+1} - (4+\ell+\ell^2+2 \ell n + 2 n^2) b_{n,\ell} = 0.\end{split}\end{equation}
		\\
		
		$\bullet$ {\bf Perturbed model:} From \crefrange{alpha-functional-sum-rule-WF1}{alpha-functional-sum-rule-WF4} we see that we require the recursion relations for $\Delta_E = \Delta_{{0,0}} + \epsilon \gamma_{0,0}^{(1)}$. As one can see from (\ref{expression-of RSTUW}), expressions of ${\mathcal R}$, ${\mathcal S}$, ${\mathcal T}$	and ${\mathcal U}$ do not depend $\Delta_E$. 
		However in this case, 	
		\be 
		\mathcal{W}_{n,\ell}= \ell + \ell^2 + 2 \ell n + 2 n^2, \quad\quad \mathcal{W}^\prime_{n,\ell}= \ell +  2 n.
		\ee
	
		Consider $\Delta_E = \Delta_{{0,0}} + \epsilon \gamma_{0,0}^{(1)}$. Once again we set $\frac{\tilde{c}_{n,\ell}}{A}=0$ and $\frac{\tilde{a}_{n,\ell}}{A}=0$ for the same reason. From (\ref{Witten-diagram}) we get 
		\be \left( \frac{B_{n,\ell}}{A} \right)_{\Delta_E = \Delta_{0,0} + \epsilon \gamma_{0,0}^{(1)} } = \alpha_{n,\ell}^{(s)} [ G^{(t)}_{\Delta_E,0}  ]_{\Delta_E = \Delta_{{0,0}} + \epsilon \gamma_{0,0}^{(1)}} \ee
		Expanding both sides and keeping only the term upto zeroth order in $\epsilon$ we can write
		\be \left(\frac{B^{(0)}_{n,\ell}}{A}\right)_{\Delta_E = \Delta_{0,0} + \epsilon \gamma_{0,0}^{(1)}} = \alpha_{n,\ell}^{(s)}[G^{(t)}_{\Delta_E,0} ]_{\Delta_E = \Delta_{0,0}}  \ee
		Referring to the duality relation of the functionals (\ref{functional-duality-property1}) and the fact that we will consider the contribution upto zeroth order of $\epsilon$ and $\Delta_{0,0}$ is the weight of a double trace operator,  we conclude $\left(\frac{B^{(0)}_{n,\ell}}{A}\right)_{\Delta_E = \Delta_{0,0} + \epsilon \gamma_{0,0}^{(1)}} =0$. A similar argument would lead to $\left(\frac{C^{(0)}_{n,\ell}}{A}\right)_{\Delta_E = \Delta_{0,0} + \epsilon \gamma_{0,0}^{(1)}} =0$. Substituting these in the recursion relation upto zeroth order we observe that it is identically satisfied.
		
		We will determine the corrections upto first order in $\epsilon$, given by  $\left(\frac{B^{(1)}_{n,\ell}}{A}\right)_{\Delta_E = \Delta_{0,0} + \epsilon \gamma_{0,0}^{(1)}}$ and $\left(\frac{C^{(1)}_{n,\ell}}{A}\right)_{\Delta_E = \Delta_{0,0} + \epsilon \gamma_{0,0}^{(1)}}$ from the recursion relation. Considering upto first order we get
		the recursion relation  is given by
		\be\begin{split}\label{recursion}
			& \ell \mathcal{R}^{  (0)}_{n+1,\ell-1} C^{(1)}_{n+1,\ell-1} + (\ell+2)  \mathcal{S}^{ (0)}_{n , \ell+1} C^{(1)}_{n , \ell+1} + \ell \mathcal{T}^{ (0)}_{ n, \ell-1} C^{(1)}_{n , \ell-1} + (\ell+2) \mathcal{U}^{ (0)}_{n-1,\ell+1}C^{(1)}_{n-1,\ell+1} + (\ell+1) \mathcal{W}^{ (0)}_{n ,\ell} C^{(1)}_{n ,\ell} = \tilde{c}_{n,\ell}
			\\
			& \ell \mathcal{R}^{(0)}_{n+1,\ell-1} B^{(1)}_{n+1,\ell-1} + (\ell+2)  \mathcal{S}^{(0)}_{n , \ell+1} B^{(1)}_{n , \ell+1} + \ell \mathcal{T}^{(0)}_{ n, \ell-1} B^{(1)}_{n , \ell-1} + (\ell+2) \mathcal {U}^{(0)}_{n-1,\ell+1}B^{(1)}_{n-1,\ell+1} + (\ell+1) \mathcal{W}^{(0)}_{n ,\ell} B^{(1)}_{n ,\ell} 
			\\
			& +  \ell \mathcal{R}^{ \prime (0)}_{n+1,\ell-1} C^{(1)}_{n+1,\ell-1} + (\ell+2)  \mathcal{S}^{\prime (0)}_{n , \ell+1} C^{(1)}_{n , \ell+1} + \ell \mathcal{T}^{\prime (0)}_{ n, \ell-1} C^{(1)}_{n , \ell-1} + (\ell+2) \mathcal{U}^{\prime (0)}_{n-1,\ell+1}C^{(1)}_{n-1,\ell+1} + (\ell+1) 
			\mathcal{W}^{\prime (0)}_{n ,\ell} C^{(1)}_{n ,\ell}  = \tilde{a}_{n,\ell}
		\end{split}
		\ee
		
		Setting $\frac{\tilde{c}_{n,\ell}}{A}=0$, we consider the recursion relation of $C^{(1)}_{n,\ell}$.  In our case  since we have already established  that $C^{(0)}_{n,\ell} = 0$ for $\Delta_E=\Delta_{0,0} + \epsilon \gamma_{0,0}^{(1)}$ by using duality relations, only the first order $C^{(1)}_{n,\ell}$ have appeared in the recurrence relation.  Upto first order the relation turns out be
		\be\begin{split}
			\frac{(\ell+1)}{A} \left(   -2 (n+1)^2 C^{(1)}_{n+1,\ell-1} - 2 (n+\ell+2)^2 C^{(1)}_{n,\ell+1} - \frac{(n+\ell-1)^2(n+\ell)^2}{2(2n + 2\ell -1)} C^{(1)}_{n,\ell-1} \right. \\
			\left.  - \frac{(n-2)^2(n-1)^2}{2(2n-3)(2n-1)} C^{(1)}_{n-1,\ell+1} + (2n^2 + 2n\ell +\ell^2+\ell)C^{(1)}_{n,\ell} \right)= 0
		\end{split}\ee
		In order to determine $C^{(1)}_{n,\ell}$ from this recursion relation we require seed values. 
		$\frac{C_{0,\ell}}{A}$ for general values of parameters is given  in \cite{mazavc2021basis} and a close examination of that with  $\Delta_E=\Delta_{0,0} + \epsilon \gamma_{0,0}^{(1)}$ reveals that  $\frac{C^{(1)}_{0,\ell}}{A} = \frac{C^{(1)}_{0,0}}{A}  \delta_{\ell,0}$. Or in other words $\frac{C^{(1)}_{0,\ell}}{A}$ vanishes for non-zero $\ell$. Using this in the recursion relation it is straightforward to show that $\frac{C^{(1)}_{n,\ell}}{A}$ vanishes for non-zero $n$ as well and we can write $\frac{C^{(1)}_{n,\ell}}{A} = \frac{C^{(1)}_{0,0}}{A}\delta_{n,0}\delta_{\ell,0}$ .
		
		Next we consider the second equation of (\ref{recursion}) and set $\frac{\tilde{a}_{n,\ell}}{A}=0$. Since $\frac{C^{(1)}_{n,\ell}}{A}$ contributes only for $n=0$ and $\ell=0$ and since the zeroth order contribution  of  $\mathcal{T}^{\prime }_{ 0,0 }$ and $ \mathcal{W}^{\prime }_{0 ,0}$ vanish the terms involving $C^{(1)}$ will decouple from the equation and we will be left with a recursion relation with $\frac{B^{(1)}_{n,\ell}}{A}$ only. A close examination of (\ref{alpha-functional-sum-rule-WF2}) shows that all the terms except  $\frac{B^{(1)}_{m,k}}{A}$ vanish for $m\neq 0$ which implies  that $\frac{B^{(1)}_{m,k}}{A}$  also has a similar behavior. Incorporating this fact in the recursion relation, we find that it gets simplified into
		\ben
		-2(\ell+2)^2  \frac{B^{(1)}_{0,\ell+1}}{A} - \frac{\ell^2(\ell-1)^2}{2(2\ell-1)(2\ell+1)}  \frac{B^{(1)}_{0,\ell-1}}{A} + \ell(\ell+1)  \frac{B^{(1)}_{0,\ell}}{A} = 0
		\ee
		Solving this recursion relation one can show 
		\be 
		\frac{B^{(1)}_{n,\ell}}{A} =  \frac{B^{(1)}_{0,0}}{A} \delta_{n,0}\delta_{\ell,0}.
		\ee
		
		\bibliographystyle{unsrt}
		\bibliography{References}
		
	\end{document}